%% file: main.tex
\documentclass[%
 reprint,
superscriptaddress,
preprintnumbers,
nofootinbib,
 amsmath,amssymb,
 aps,
 prd,
floatfix,
]{revtex4-2}

\usepackage{graphicx, subcaption}
\usepackage{dcolumn}
\usepackage{bm}

\usepackage{cancel}
\usepackage{soul}
\usepackage[normalem]{ulem}
\usepackage{siunitx}

\usepackage{hyperref}
\hypersetup{
    colorlinks=true
}

\begin{document}

\title{Macroscopic approach to the radar echo scatter from high-energy particle cascades}

\input{author_list}

\date{\today}

\begin{abstract}
To probe the cosmic particle flux at the highest energies, large volumes of dense material like ice have to be monitored. This can be achieved by exploiting the radio signal. In this work, we provide a macroscopic model to predict the radar echo signatures found when a radio signal is reflected from a cosmic-ray or neutrino-induced particle cascade propagating in a dense medium like ice. Its macroscopic nature allows for an energy independent run-time, taking less than 10~s for simulating a single scatter event. As a first application, we discuss basic signal properties and simulate the expected signal for the T-576 beam-test experiment at the Stanford Linear Accelerator Center. We find good signal strength agreement with the only observed radar echo from a high-energy particle cascade to date.
\end{abstract}

\maketitle

\section{\label{sec:introduction}Introduction}
The discovery of the cosmic neutrino flux by the IceCube collaboration in 2013 established neutrino astronomy as a new window on the Universe~\cite{IceCube2013}. The IceCube experiment covers a volume of roughly 1 km$^3$ instrumented with optical Cherenkov detectors to probe the remnants of the neutrino-ice interaction. The instrumentation of this volume allows the detection of cosmic neutrinos in the TeV-PeV energy range, but due to the steeply falling neutrino flux, IceCube runs low in statistics above PeV energies~\cite{IceCube2021uhz}. As such, detecting the cosmic neutrino flux above PeV energies, requires probing larger volumes than the 1 km$^3$ that is currently monitored by IceCube. 

Radio waves travel roughly 10 times further than optical signals in ice~\cite{Besson2008,Aguilar2022}, and are therefore a promising probe for instrumenting such larger detection volumes. Many facilities~\cite{Schroder2017} are based on passively detecting coherent radio Cherenkov emission from electromagnetic showers, namely the Askaryan effect~\cite{Askaryan1962,T5102022,CODALEMA2015, PierreAuger:2014ldh}. Here, we discuss {\it active} radio detection of high energy particle cascades in dense media through the radar echo technique. A radio transmitter constantly illuminates a volume with radio waves. When a particle cascade appears in the volume, the radio waves reflect off the plasma left in its wake, and are observed with a radio receiver. Here it should be noted that the word {\it plasma} has particular interpretations in various fields. In this paper, we consistently use the word to refer to the medium created by the particle cascade. A significant advantage of the radar echo method is that the user has full control over the transmitted radio signal's amplitude and frequency content, and therefore control on the expected signal upon reflection. In much the same way that signal-to-noise is greatly enhanced in radio communication systems, the radar echo method offers great promise to detect high energy events.
 
The earliest attempts to use the radar echo method to probe high-energy particle cascades date from the 1940s, when Blackett and Lovell theorized that the method may be suitable for the in-air detection of cosmic-ray air showers \cite{Blackett1941}. Concurrently, in 1941, Eckersley showed that a crucial component in earlier radar scatter modeling efforts was missing: the collisional damping of the free ionization charges responsible for the possible scatter \cite{Lovell1993}. As will be shown in Sec. II, the main effect of this is that the plasma constituents, the free ionization charges, need to be treated as a collection of individual scattering objects, rather than a perfect coherent reflector over the plasma surface. 

In spite of the calculation by Eckersley, several attempts were still made over the years to probe cosmic-ray air showers~\cite{Lovell1993, Matano1968}, with a revival of the method in the early 2000s~\cite{Gorham2001}. However, further theoretical advancements~\cite{Stasielak2013, Filonenko2013}, as well as limits obtained by the TARA experiment~\cite{TARA2017}, finally deemed the radar echo method insufficient to probe cosmic-ray air showers.  

Around the same time (2010-2015), it was realized that cascades in dense media provide much more favorable detection properties. The main advantage is found in the drastic increase of the ionization plasma density left in the wake of the cascade~\cite{chiba2012,deVries2015}. This triggered several experimental efforts to probe particle cascades propagating in dense media in the laboratory~\cite{devries2015b,Prohira2019b}, leading to the successful radar detection of such a cascade at the Stanford Linear Accelerator Center (SLAC) T-576 experiment~\cite{Prohira2020} by the Radar Echo Telescope (RET) collaboration. 

This successful detection initiated the radar echo telescope for cosmic rays (RET-CR) experiment. Its goal is to probe the in-ice continuation of cosmic-ray air shower cores penetrating a high altitude ice sheet~\cite{deKockere2022,RETCR2021}. RET-CR uses a cosmic-ray surface detector located on top of the ice sheet to trigger an in-ice radar detector. The in-ice detector aims to detect the continuation of the cosmic-ray particle cascade into the ice. As such, detecting this signal will show the in-nature proof of concept of the radar echo method to probe in-ice particle cascades. After the RET-CR experiment, the final goal of the RET collaboration is to perform neutrino astronomy using the radar echo telescope for neutrinos (RET-N)~\cite{RadarEchoTelescope:2021wph}. 

Although several modeling efforts for scattering radio waves off of in-air particle cascades have been developed~\cite{Gorham2001, Takai2011, Stasielak2013, Filonenko2013}, currently the only available model to simulate the radar echo process from relativistic particle cascades in dense media is \textsc{RadioScatter}~\cite{Prohira2019}. \textsc{RadioScatter} provides a particle level simulation of scattering from ionization deposits, which can be generated by, for example, the \textsc{GEANT4}~\cite{geant4} toolkit. The nominal \textsc{RadioScatter} setup presented in~\cite{Prohira2019} simulates a particle cascade with a low-energy primary and, if necessary,  uses an energy scaling and cascade profile stretching to provide the results for the user specified target energy, which can be different from the initial primary particle energy. The \textsc{RadioScatter} simulation run-time, when using \textsc{GEANT4} as the cascade generator, scales with energy, ranging from $O$(1-10~s) for a 10 GeV primary up to $O(1000$~s) for a 10 TeV primary. \textsc{RadioScatter}'s architecture results in a trade-off between run-time (faster for lower energy primaries) and resolution (better for higher energy primaries).

In this work, we present a complementary, deterministic, modeling framework called \textsc{MARES}, a macroscopic approach to the radar echo scatter. \textsc{MARES} provides an analytic, macroscopic treatment of the radar echo problem, using cascade profile parametrizations to obtain the ionization profile. This approach results in an energy independent run-time of less than 10~s for a single scatter event, resulting in significant speedups for large-scale simulations. The simulation code is written in C++, and will be made publicly available.

We first outline the scattering formalism by introducing the radar cross section (RCS) for a macroscopic volume of coherent scatterers. Subsequently, the approach is extended to cover the full cascade, including its relativistic propagation, the latter leading to a nontrivial cascade RCS. As a first application, we discuss basic signal scattering properties and apply our model to the T-576 experiment at SLAC, showing good agreement with the observed signal.   

\section{\label{sec:scatter}Radar scattering from a macroscopic volume of charges}

The object we intend to detect is the ionization trail left in the wake of a relativistically moving particle cascade. Therefore, it is crucial to work at the electric field level, such that we are able to trace the phase information and determine the interference between different emission points within the plasma. As such, starting from the radar range equation, we first derive its electric field equivalent to obtain the field amplitude at the scattering point, which is subsequently reemitted to the receiver. The signal phase and amplitude are traced independently.

The base of our calculation is initially found in the radar range equation \cite{BalanisAntenna}. This allows us to predict the scatter from a static macroscopic target having a radar scatter cross section $\sigma_{RCS}$,
 \begin{equation} \label{eq: radar_power_2}
     P_R = P_T \frac{ G_T G_{R} \lambda^2 }{(4\pi)^3 (R_T R_R)^2 } \sigma_{RCS},
 \end{equation}
where the received power, $P_R$, is a function of the transmitted power, $P_T$; the gain of the transmitter and receiver antennas, $G_T$ and $G_R$; the distances of both antennas to the scattering element, $R_T, R_R$; and the wavelength, $\lambda$, of the radio wave. 

Using the medium impedance, for this work taken to be ice, $Z_{ice}$, and the antenna effective area $A_R=\left(\frac{\lambda^2 G_R}{4\pi}\right)$ ~\cite{BalanisAntenna}, the received power is related to the electric field amplitude $|E_R|$,  

\begin{eqnarray} \label{eq:power-field_relation}
    P_{R}&& = \frac{ E_R^2}{2Z_{ice}} A_{R}, \\
    |E_R|&& =\sqrt{ 2Z_{ice}P_R \left(\frac{\ 4\pi}{\lambda^2 G_R}\right) }. 
\end{eqnarray}
This now allows us to rewrite Eq.~(\ref{eq: radar_power_2}) in terms of the received electric field amplitude,
\begin{equation} \label{eq: radar_field}
    |E_R| = \frac{1}{4\pi R_TR_R}\sqrt{2Z_{ice}P_TG_T}\sqrt{\sigma_{RCS}}.
\end{equation}

\subsection{The damped electron radar cross section}
The scattering object under consideration in this work is the ionization plasma left in the wake of a high-energy particle cascade in ice. The calculations reported here are based on a model of the medium dominated by collisional damping of the free charges that are driven by an incoming electromagnetic field. The collisional damping originates from elastic collisions with (neutral) medium particles. Based on collision rates of free electrons in ice, as derived in~\cite{Prohira2019} that is consistent with the electron mobility results found in \cite{DeHaas1983}, a collisional damping rate parameter of $\nu_c \approx 100\; \pm\; 50$~THz for free electrons in ice is taken as the base value in our model.
In our model, $\nu_c$ is much larger than the typical radio frequencies in the $\nu=\frac{\omega}{2\pi}\sim$~MHz$-$GHz range at which we aim to probe the plasma, as well as the characteristic plasma frequency, 
\begin{equation}
    \omega_p~[\mathrm{{rad}/s}]~
    =\left( \frac{n_eq^2}{m\epsilon_0} \right)^{1/2}; ~
    \nu_p = \approx8980\sqrt{n_e\;[\mathrm{{e^-}/cm^3}]},
\end{equation}
which is also found in the GHz range.

Given that the collision frequency is the leading scale in the system, the plasma is treated as a collection of individual scattering objects, oscillating in the incoming field, with their movement strongly damped by collisions with the medium particles. 

An emitter with transmit power $P_T$ and gain $G_T$ induces an electric field at the single charge equal to,
\begin{equation} \label{eq: E_field_at_e}
    |E_e|=\sqrt{\frac{2Z_{ice}P_TG_T}{4\pi R_T^2}}.
\end{equation}

In the overdamped regime, the equation of motion for a single free charge that is driven by the electromagnetic force is properly obtained using a damped-driven oscillator model. As detailed in Appendix~\ref{app:eom}, using this model to obtain the charge movement, the field at the receiver is subsequently found through the standard field equations and can be expressed equivalent to Eq.~(\ref{eq: radar_field}),
\begin{equation} \label{eq: field-e-R}
\vec{E}_{e,R} =  \frac{1}{4\pi R_TR_R}\sqrt{2Z_{ice}P_TG_T}\sqrt{\sigma_{RCS,e}},
\end{equation} 
defining the radar scattering cross section of the damped free electron as
\begin{equation}\label{eq: RCSe}
    \sigma_{RCS,e} = (\omega^2W)^2 \sigma_{Th} G_{Hz}. 
\end{equation}
The cross section is written in terms of the Thomson scattering cross section $\sigma_{Th} = \frac{8\pi}{3}r_e^2$, for which all constants sum to the classical electron radius $r_e~\sim~2.8~\times~10^{-15}$~m, and the Hertzian dipole gain factor $G_{Hz}~=\frac{3}{2}\sin^2(\theta)$, where $\theta$ denotes the opening angle between the oscillation direction and the line of sight toward the observer. Finally, an additional damping term $\omega^2 W$ is obtained, with $W$ being 
a function of the characteristic frequencies in the system, defined as
\begin{equation}
W=\left(\frac{1}{\omega^2 - i\omega \nu_c} \right). 
\end{equation} 

\subsection{Scattering from a macroscopic volume}

\begin{figure} 
     \centering
     \includegraphics[width=\linewidth]{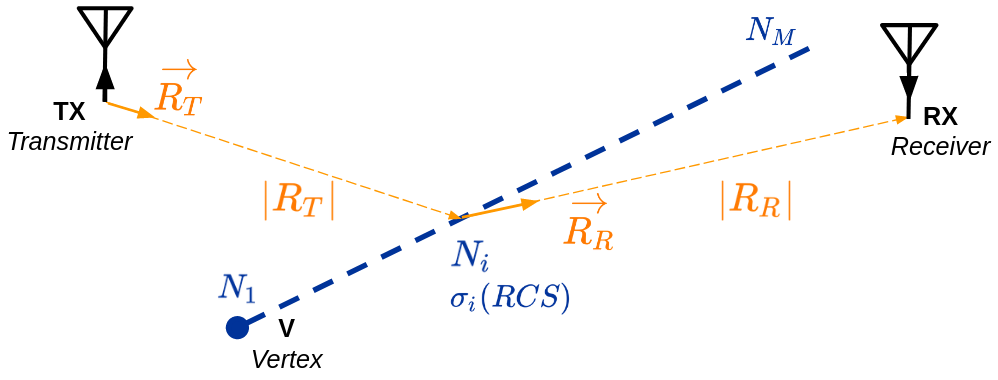}
     \caption{Basic diagram of the radar scatter of a line of scattering centers (segments).}
     \label{fig:scatter_sketch}
\end{figure}

In this work, we construct our model based on the scatter off of $M$ coherent segments each containing a collection of $N_i$, $i=1,...,M$, scatterers. The geometry of this setup is outlined in Fig.~\ref{fig:scatter_sketch}. In order for a segment to satisfy the coherence condition, the dimensions of its scattering volume $d_x,d_y,d_z$ have to be much smaller than the relevant dimensions in the system. The relevant dimensions are given by the transmitter-cascade distance, $R_{T}$; the cascade-receiver distance, $R_{R}$ and the wavelength at which the plasma is probed, $\lambda_T$. As such, we require $d_x,d_y,d_z \ll R_T,R_R,\lambda_T$. Using this condition, the electric fields of the $N_i$ individual electrons within a segment will add constructively at our receiver. 

The received power under coherence scales as $N_i^2$, hence the scatter from a single segment is given by
\begin{equation} \label{eq: Efield-linearity}
    P_{R,i} \propto |\vec{E}_{R,i}(R_i,t)|^2 =
    |N_i {E}_{e,R_i}(R_i,t)|^2,
\end{equation}
with $E_{e,R_i}$ the scattered field from a single electron given by Eq.~(\ref{eq: field-e-R}).

\subsection{Energy balance} \label{app:energy_balance}

For cold, collisionless electrons undergoing Thomson scattering, the scattering process is elastic, and the energy lost by the wave crossing the plasma is exactly equal to the energy in the radar echo. 
In this section, we show that the energy balance can be obtained from the microscopic perspective outlined in the previous section, as well as from a macroscopic perspective considering the change in the refractive index in the medium induced by the particle cascade. 

From a macroscopic perspective, following \cite{Prohira2019, JacksonClassical}, in the plasma under consideration, $\mu=\mu_0$ and $\epsilon=\epsilon_0(1+\chi_e)$, the dispersion relation for the propagation of an incoming electromagnetic wave becomes, to good approximation \cite{JacksonClassical},
%
 \begin{eqnarray} \label{eq: dispersion_relation}
    k&&=\frac{\omega}{c}\left[ 1+\omega_p^2 W \right]^{1/2} \simeq \frac{\omega}{c}\left[ 1+\frac{1}{2}\omega_p^2 W \right]\nonumber\\
    &&= \mathrm{Re}(k) + i\; \mathrm{Im}(k),
 \end{eqnarray}
The propagating radio wave ($\vec{k}$) gains an imaginary term, the plasma attenuation, $\beta$:
\begin{eqnarray}
    \beta &&= 2\; \mathrm{Im}(k) = 2 \frac{\omega}{c}\; \mathrm{Im}(n) \simeq \frac{\omega}{c}\omega_p^2\; \mathrm{Im}(W) \nonumber\\
    &&= \frac{\nu_c}{c} (\omega \omega_p W)^2. 
\end{eqnarray}
The integrated attenuation of the radio waves considered along a path is the radio transparency, $\mathcal{T}$:
\begin{equation}\label{eq:transparency}
    P(z) = P_0 e ^{\int_0^z -\beta(l) dl } = P_0 \mathcal{T}.
\end{equation}
The attenuation of a radio wave going through a plasma volume $V$ of thickness $l$ and area $A$ can now be obtained as
\begin{eqnarray}
\mathcal{P}_{att} &&= \frac{\Delta P}{V} = \frac{1}{V}P_{0}(1 - \mathcal{T}) \nonumber\\
 &&= \frac{IA}{Al} (1 - e^{-\beta l}) \approx \frac{I}{l} \beta l = I\beta,
\end{eqnarray}
where $I=\frac{\epsilon_0 c}{2} E_0^2$ defines the irradiance, and in the relevant overdamped regime, $\omega, \omega_p \ll \nu_c$, $\beta \to\frac{\omega_p^2}{c\ \nu_c}$ becomes small.

 Since we previously obtained the radar scatter from a microscopic, single particle approach, we can compare the macroscopic derivation above to the attenuation of the incoming wave due to the radar scatter calculated at the individual particle level. The energy taken from the incoming wave by a single electron and which is rescattered is~\cite{TaylorClassical}
\begin{equation} \label{eq: DDO-power}
    \left< P(t) \right> = \left< {F}(t) \dot{x}_e \right> = \frac{1}{2m} (q E_0 \omega W )^2 \nu_c.
\end{equation}
For a volume $V$ containing $N$ electrons with electron number density $n_e$, the power density $ \mathcal{P} = \frac{P}{V} = \left< P(t) \right> n_e $ is 
\begin{equation}
    \mathcal{P}_{loss} = \frac{1}{2} \epsilon_0 (\omega_p E_0 \omega W )^2 \nu_c \\ 
\end{equation}
Using the electron energy loss as the source of attenuation of the incoming radio wave, $\mathcal{P}_{loss}=I \beta$, and the irradiance $I=\frac{\epsilon_0 c}{2} E_0^2$, we can now recover the same attenuation $\beta=\frac{\nu_c}{c} (\omega \omega_p W)^2$ that we have defined through pure macroscopic quantities.

It follows that our plasma is indeed well described by the damped driven oscillator model, and the energy balance is obtained from both microscopic and macroscopic considerations. This also implies that the plasma needs to be treated as a collection of individual scattering objects oscillating and reemitting in the incoming field, contrary to the so-called overdense regime ($\omega, \nu_c \ll \omega_p$), where the plasma would act as a perfect reflector scattering over its macroscopic surface. 

\section{The particle cascade}\label{sec:cascade}
We now apply the formalism developed in the previous section to the ionization plasma left in the wake of our particle cascade. To do so, we need a model of our plasma. For this, we first consider the relativistic particle cascade itself. Its longitudinal profile is parametrized as a function of column depth $X[\mathrm{g/cm}^2]$ following Greisen~\cite{GreisenCR, Matthews2010}:
\begin{equation} \label{eq:Greisen}
N(X) = \frac{0.31}{\sqrt{\text{ln}(E/E_{crit})}} e^{\frac{X}{X_0 (1-1.5\text{ln}(s))}}, 
\end{equation}
with $E[\mathrm{MeV}]$ the energy of the primary cascade-inducing particle, and $s$ the shower age. The lateral particle distribution inside the cascade front is taken from the Nishimura and Kamata parametrization~\cite{Kamata1958}, 
\begin{equation} \label{eq:NishimuraKamata}
    w(r,s) = \frac{\Gamma(4.5 -s)}{\Gamma(s) \Gamma(4.5 - 2s)} \left(\frac{r}{r_0}\right)^{s-1} \left(\frac{r}{r_0} + 1 \right)^{s-4.5}.
\end{equation}
The parameters in these models can easily be adapted for ice instead of air by using the critical energy $E_{crit} = 78.6$ MeV; the Molière radius $r_0 = 7$ cm; and the interaction length $X_0(ice) = 36.08\; \text{g}/\text{cm}^2 \simeq X_0(air)$~\cite{deVries2015}. Given the longitudinal particle spread in the cascade front is much smaller than the radial spread, this is safely approximated by the delta function $w(r,h)=w(r)\delta(h)$. From this, the number of ionization particles within a longitudinal step of $dX=1\;\mathrm{g/cm^2}$, corresponding to a length $dl=dX/\rho_{ice}$~cm is obtained assuming a constant average ionization loss of 2~$\mathrm{MeV/g/cm^2}$. 

\begin{figure} 
     \centering
     \includegraphics[width=\linewidth]{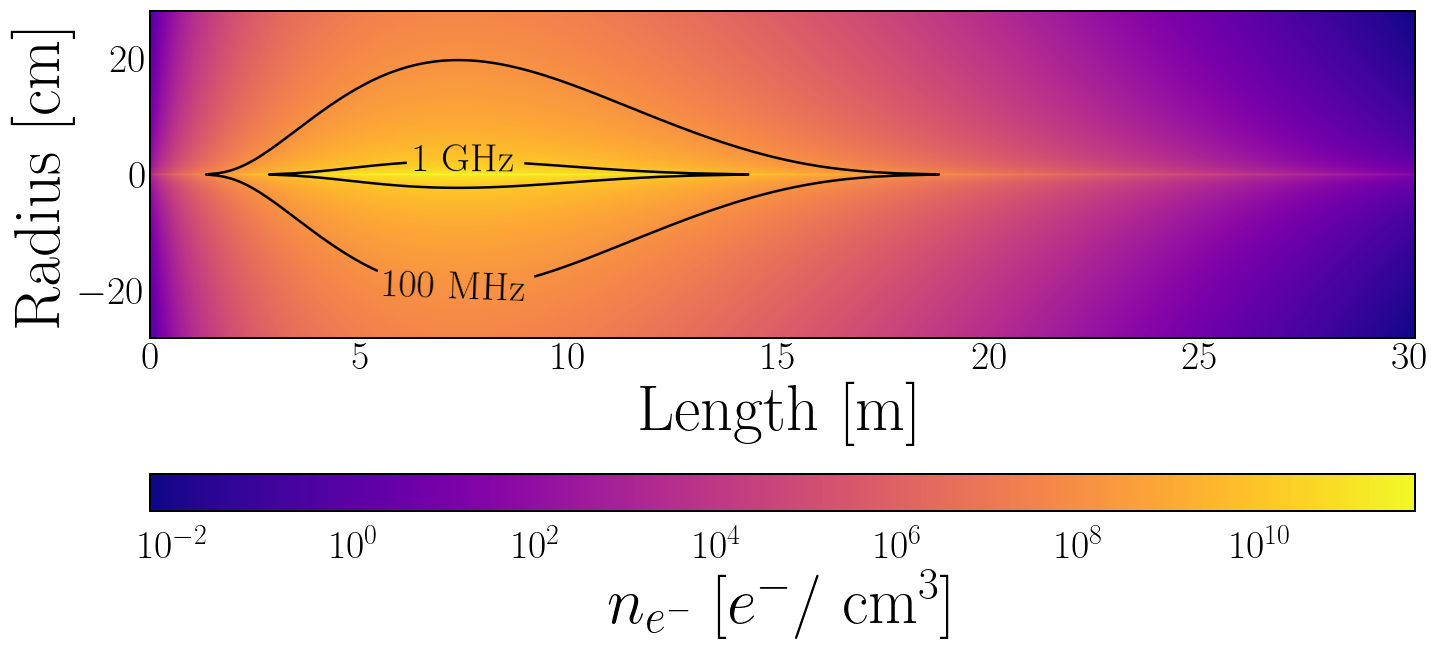}
     \caption{The cascade density profile $n_e$ of a $10^7$ GeV primary. The density and plasma frequency $\nu_p $ scale with the energy of the cascade's primary, $E_p$. Curves of constant plasma frequency (and density) have been added for reference.}
     \label{fig:density_fig}
\end{figure}

As we are interested in the number of ionization electrons, we have to consider a mean ionization energy. Following~\cite{Timneau2004}, this is chosen at 20~eV. Assuming an ionization energy of 20~eV, the number of ionization electrons over a length $dX$ becomes $N_{ion,e}dX=10^5$ per relativistic lepton inside the particle cascade. The free ionization charge density at a depth $X$ within a volume $dV(r)=2\pi r \;dr \;dl$ is now obtained as
\begin{equation}
    n_{e, plasma}(X,r) \simeq  \frac{N_{ion,e}dX}{dV} \int_r^{r+dr} N(X,E_p) w(r',s)dr'.
\end{equation}
The obtained density profile for a $10$~PeV particle cascade is shown in Fig.~\ref{fig:density_fig}.

It is important to note that the number of particles that are instantly available for scattering depends strongly on the lifetime of the free electron plasma left in the wake of a high energy particle cascade. This lifetime is found to be of the same order as other relevant timescales in the system. The total cascade propagation time is $t_p\sim O$(10-100)~ns, and the probing frequency has a period of $t_f\sim O$(1-20)~ns. The plasma lifetime, $\tau$, is a function of the medium temperature and purity~\cite{DeHaas1983}, and found to be of the order $\tau\sim O$(1-50)~ns (see Appendix~\ref{app:fel}). The effect of different plasma lifetimes is discussed in Sec.~\ref{sec:lifetime}, and unless otherwise indicated, for now, we will consider a lifetime of $\tau=10$~ns throughout this work. 

\section{Scattering off of the cascade} \label{section:geometry}

In the tens of MHz to GHz regime, the transmitted wavelength, $\lambda_T$, becomes the smallest scale under consideration, $O$(10 cm-10 m). Therefore, we cannot approximate the ionization plasma left in the wake of a high-energy particle cascade as a pointlike object, but have to consider its structure. 

A more general description is found in describing the cascade as a collection of $M$ segments, each consisting of $N_i$, $i=1,...,M$ coherently scattering charges. The segment volumes need to be small enough to preserve coherence under $\lambda_T$. As such, in this work we use volumes of size $O(0.1\;\mathrm{cm}^3)$. The volume size of an individual segment is, however, a free parameter of the model, and we have confirmed that the simulation results presented here converge under decreasing segment volumes (see Appendix~\ref{app:convergence}). 

In determining the observed field amplitude at our receiver from the full cascade, we now have to account for interference between the segments. Segmenting our cascade into $M$ coherent scattering volumes and considering the individual phases of the scattered field at the receiver allows obtaining the power at the receiver,
\begin{eqnarray} \label{eq: Efield-linearity2}
    P_{R,cascade} \propto |\vec{E}_{R,cascade}(R, t)|^2= \nonumber\\
     |\sum_{i=1}^M {E}_{R,i}(R_i,t)*e^{i(kR_i - \omega t + \psi)}|^2 .
\end{eqnarray}
The phase per segment $\phi_i=kR_i - \omega t + \psi$ is composed of three terms:
\begin{itemize}
    \item $kR_i = k(R_{T,i} + R_{R,i})$ captures the phase shift over the propagation of the wave from the transmitter scattered to the receiver.

    \item $\omega t$ captures the phase shift at the receiver, considering the time of emission. 
    
    \item $\psi$ is the unique relative phase per element. Under the damped-driven oscillator model, there is a frequency dependent phase difference between the driving force and the movement of the oscillator,
    \begin{equation}
        \psi = \arctan \left( \frac{\nu_c \omega}{\omega_0^2 - \omega^2} \right) = \arctan \left(-\frac{\nu_c}{\omega} \right).
    \end{equation}
    In the overdamped limit under consideration ($\nu_c \gg \omega$), this phase is constant and equal to $\psi \simeq -\pi/2$.
\end{itemize}
Three additions have to be made to our model when considering scattering off of a realistic cascade. 

First is the inclusion of the lifetime of the plasma under consideration. Including lifetime effects, the total number of particles inside a scattering segment becomes time dependent and is written as
\begin{equation} \label{eq:lifetime_factor}
    N_i(t,t_0)=N_{0,i}\cdot e^{-\frac{t-t_0}{\tau}} \Theta(t - t_0),
\end{equation}
with $t_0$, the time at which the segment is created, and $\Theta(x)$ denoting the Heaviside step function.

Second, we have to take into account that the incoming wave is not only attenuated by the medium, but that part of its power is scattered away by other segments along the line of sight between the transmitter and the segment under consideration as outlined in Sec.~\ref{sec:scatter}. To account for this effect, we have to include the transparency factor, ${\cal T}$, defined in Eq.~\ref{eq:transparency}. The transparency factor will be included into the scattering cross section for a single cascade segment. Therefore, the single segment scattering cross section does not stand by itself, but rather considers previous segments along the line of sight to the transmitter. Incorporating both effects, the radar scattering cross section for a single scattering segment is given by
\begin{equation}
\label{eq: full_RCS}
    \sigma_{RCS,i}(t,t_0) = \sigma_{Th} N_{e,i}^2 (\omega^2W)^2 \cdot G_{Hz}\cdot {\cal T} \cdot e^{-\frac{2(t-t_0)}{\tau}} \Theta(t - t_0).
\end{equation}

The third and final effect to include is the attenuation of the incoming wave by the medium in which it propagates before reaching the plasma. The attenuation length is a free parameter in our model and attenuates the final field by a factor,
$E_{R,i}=E_{R,i}^{no\;att}e^{-(R_{T}+R_{R})/L_{att}}$.
In the following, we fix the attenuation length to $L_{att}=1.45$~km~\cite{Besson2008}. Given that the attenuation length varies with location~\cite{Besson2008, Aguilar2022}, a specific attenuation length or parametrization is left to site-specific studies. Implementing this into Eq.~(\ref{eq: field-e-R}), now allows us to obtain the amplitude of the field from a single cascade scattering segment at the receiver as
\begin{equation} \label{eq: E_from_RCS}
|E_{R,i}|= \frac{\sqrt{2Z_{ice} P_TG_T}}{4\pi R_T R_R} \sqrt{\sigma_{RCS,i}}e^{-(R_{T}+R_{R})/L_{att}}.
\end{equation}

\subsection{Geometry}

Given the different scales in which the relevant physics of the scatter occur, different effects are evaluated in different frames of reference. The three relevant frames considered in this work are shown in Fig.~\ref{fig:frames} and discussed in detail in Appendix~\ref{app:reference}. The first frame is the bistatic scattering frame $\{\vec{e_x},\vec{e_y},\vec{e_z}\}$, defined by the transmitter and receiver locations. The cascade frame defined by $\{\vec{e_L},\vec{e_R},\vec{e_n}\}$ and the plane of incidence frame defined by $\{\vec{e_U},\vec{e_V},\vec{e_W}\}$, are both needed to compute the cascade response and the transparency factor ${\cal T}$. 

\begin{figure}[!h]
     \centering
     \begin{subfigure}[b]{\linewidth}
         \centering
         \includegraphics[width=\textwidth]{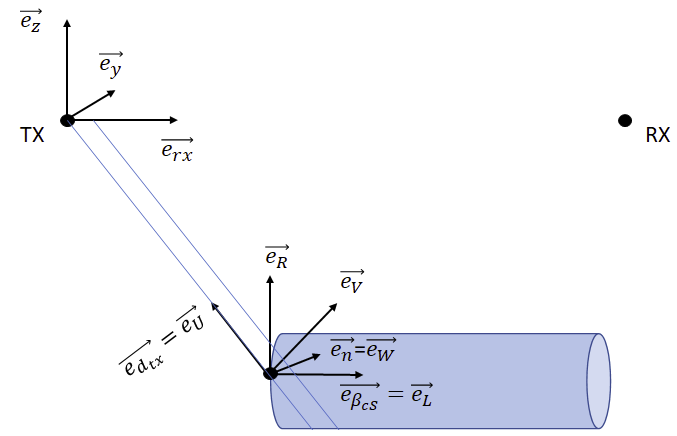}
         \caption{The reference frames considered for this work. For illustration purposes, the cascade is directed along the $\vec{e}_{RX}$ axis in the $(\vec{e}_{rx},\vec{e}_z)$ plane. In this case, the vectors $\vec{e}_y$, and $\vec{e}_n=\vec{e}_W$ are directed into the paper and all other vectors lie in the transmitter, receiver, cascade plane.}
         \label{fig:frames}
     \end{subfigure}
     \hfill
     \begin{subfigure}[b]{\linewidth}
         \centering
         \includegraphics[width=0.75\textwidth]{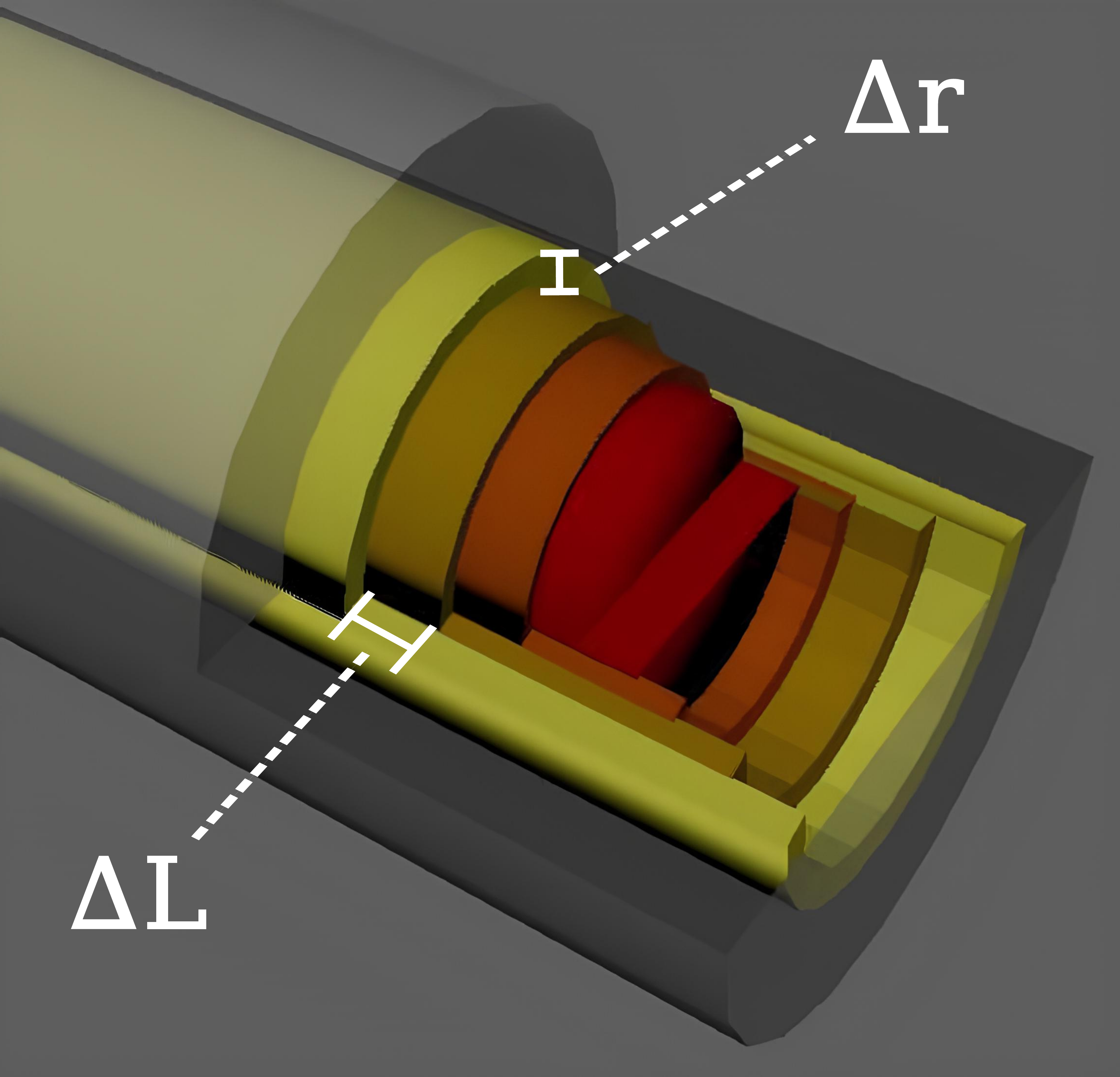}
         \caption{Illustration of the cascade segmentation into semicircular shells.}
          \label{fig:cascade_segment}
     \end{subfigure}
     \hfill
   \caption{ The segmentation of the cascade. The cascade is sliced along the line of sight vector $\vec{e}_U$ (a). The segmentation is done using semicircular shells of width $\Delta L=1$~cm, radially subdivided with a thickness of $\Delta r=1$~mm (b).}
\end{figure}

As the electron density distribution is radially symmetric, the logical segment choice would be to consider circular shells  of approximate equal density. As illustrated in Fig.~\ref{fig:cascade_segment}, in this situation, the incoming wave, however, would first traverse the top part of the shell (with thickness $\Delta r$ and width $\Delta L)$, before crossing the bottom. To properly account for the energy scattered away by the top part of the shell, before the incoming wave would reach the bottom half of the shell, the cascade segmentation is done by considering semicircular shells in the cascade frame defined along the line of sight direction $\vec{e}_U$.

Each shell has a constant radial thickness of $\Delta r=1$~mm and width along the cascade axis of $\Delta L=1$~cm. Both dimensions are chosen such that the particle density is, to very good approximation, constant within the segment, which is small enough to fulfill the coherence requirement. These parameters have been checked for convergence (see Appendix~\ref{app:convergence}), and they remain independent of the results presented here.  

A further approximation is subsequently made by positioning all semicircular shells along a ray path into a single coherent scattering center located at the intersection of the incoming ray with the cascade axis, defined by the center of the innermost shell. Placing the segments at the cascade axis along the line of sight from the transmitter has the advantage that ${\cal T}$ is directly obtained and numerically efficient to implement, but does introduce a small error in our calculation due to the slightly incorrect placement of the segments. However, given that most of the particles are located (very) close to the cascade axis, the induced error is small. For example, a ray crossing a cascade segment at $r=2$~cm from the cascade axis under 80 degrees incidence is placed incorrectly by $O(\mathrm{cm})$.   

\section{Results}
In this section, we discuss relativistic propagation effects as well as the effect of the finite lifetime of the cascade. Furthermore, we apply our model to the T-576 experiment and confirm that the observed scatter is consistent with our model prediction.

\subsection{Lifetime }
\label{sec:lifetime}

\begin{figure*}
     \centering
    \begin{subfigure}[b]{0.8\linewidth}
          \centering
     \includegraphics[width=0.8\linewidth]{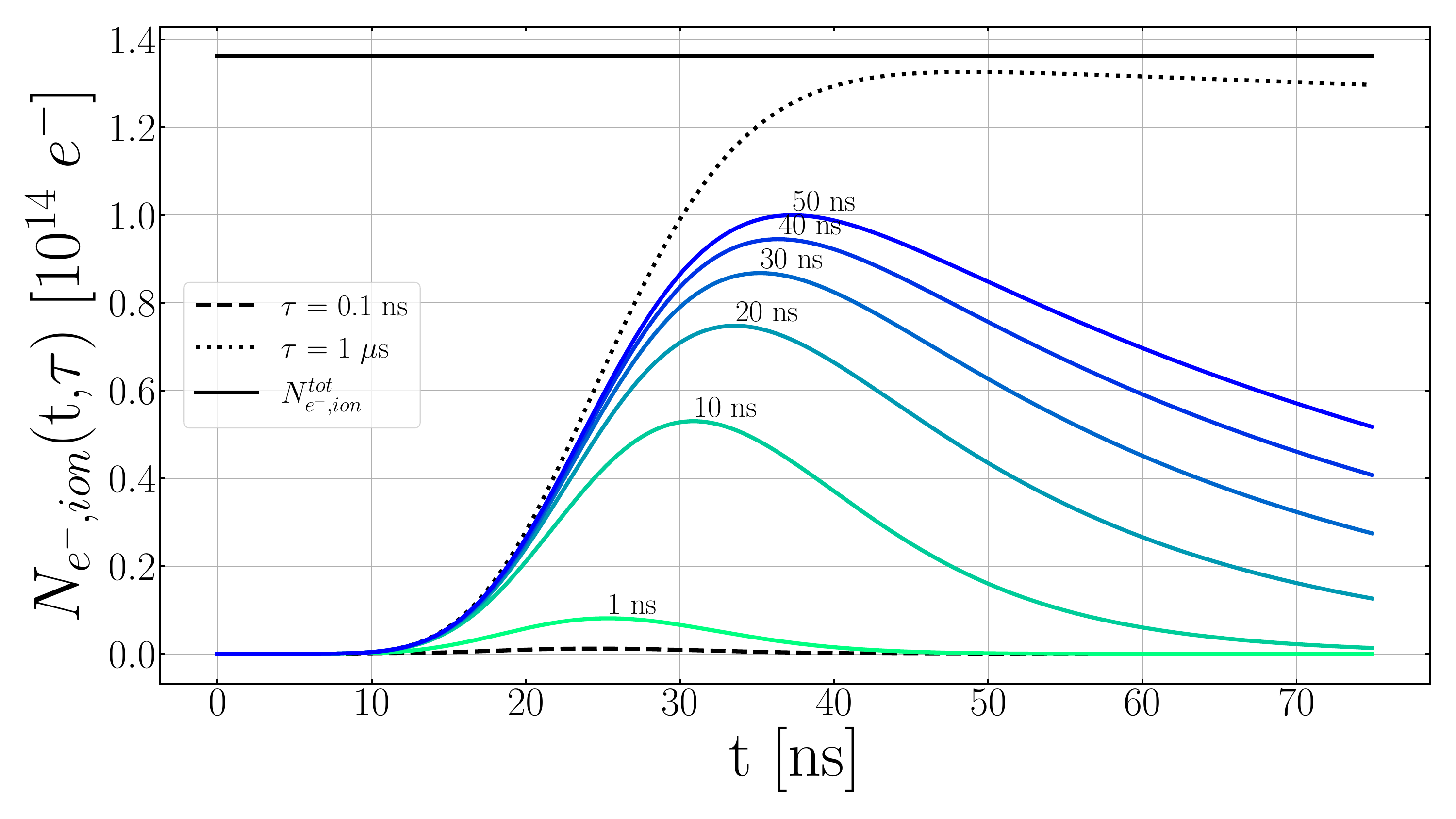}
     \caption{The free ionization electron number as function of time for a $E_p = 10^7$ GeV cascade. The straight, top line is the total amount of electrons ionized by the cascade, and represents the theoretical maximum of electrons that can be scattered off.}
      \label{fig:lifetime_10PeV}
    \end{subfigure}
    \begin{subfigure}[b]{0.8\linewidth}
          \centering
    \includegraphics[width=0.8\linewidth]{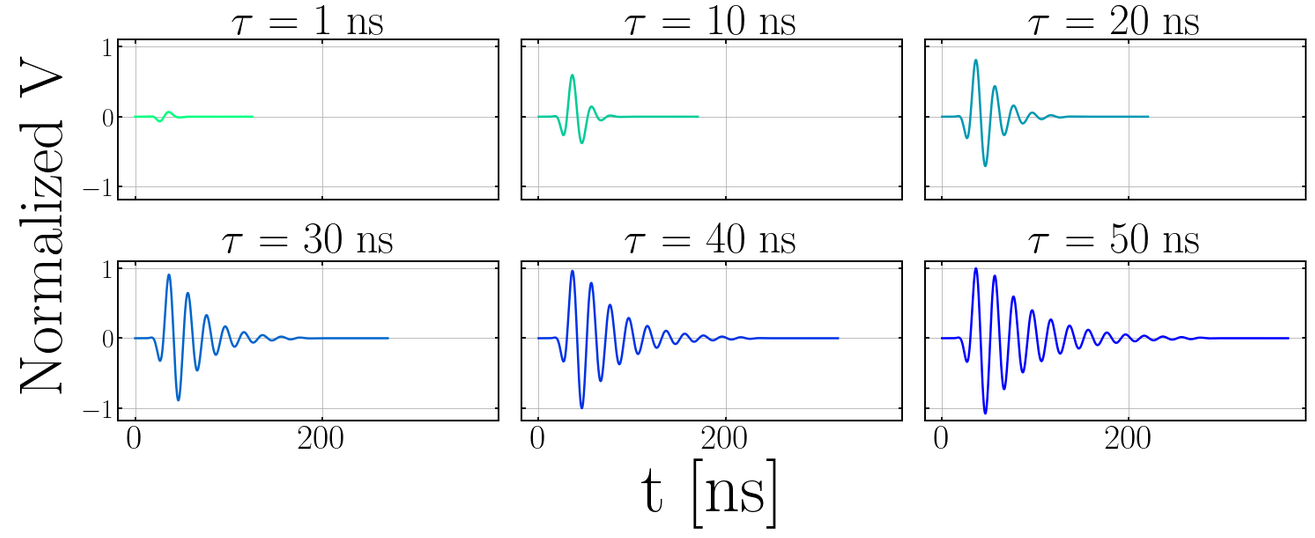}
    \caption{Illustration of the effect of different free ionization electron lifetimes on the expected pulse shape. The pulse amplitude is normalized to its peak found for $\tau=50$~ns.}
    \label{fig:lifetime_waveforms}
    \end{subfigure}
    \caption{The effect of the ionization electron plasma lifetime on the total number of particles available for scattering at any instant in time (a), and the observed (normalized) pulse shape at the detector (b).}
\end{figure*}

\begin{figure}[ht!]
     \centering
    \begin{subfigure}[b]{\linewidth}
         \centering
         \includegraphics[width=\textwidth, trim={16cm 23cm 10cm 9cm},clip]
         {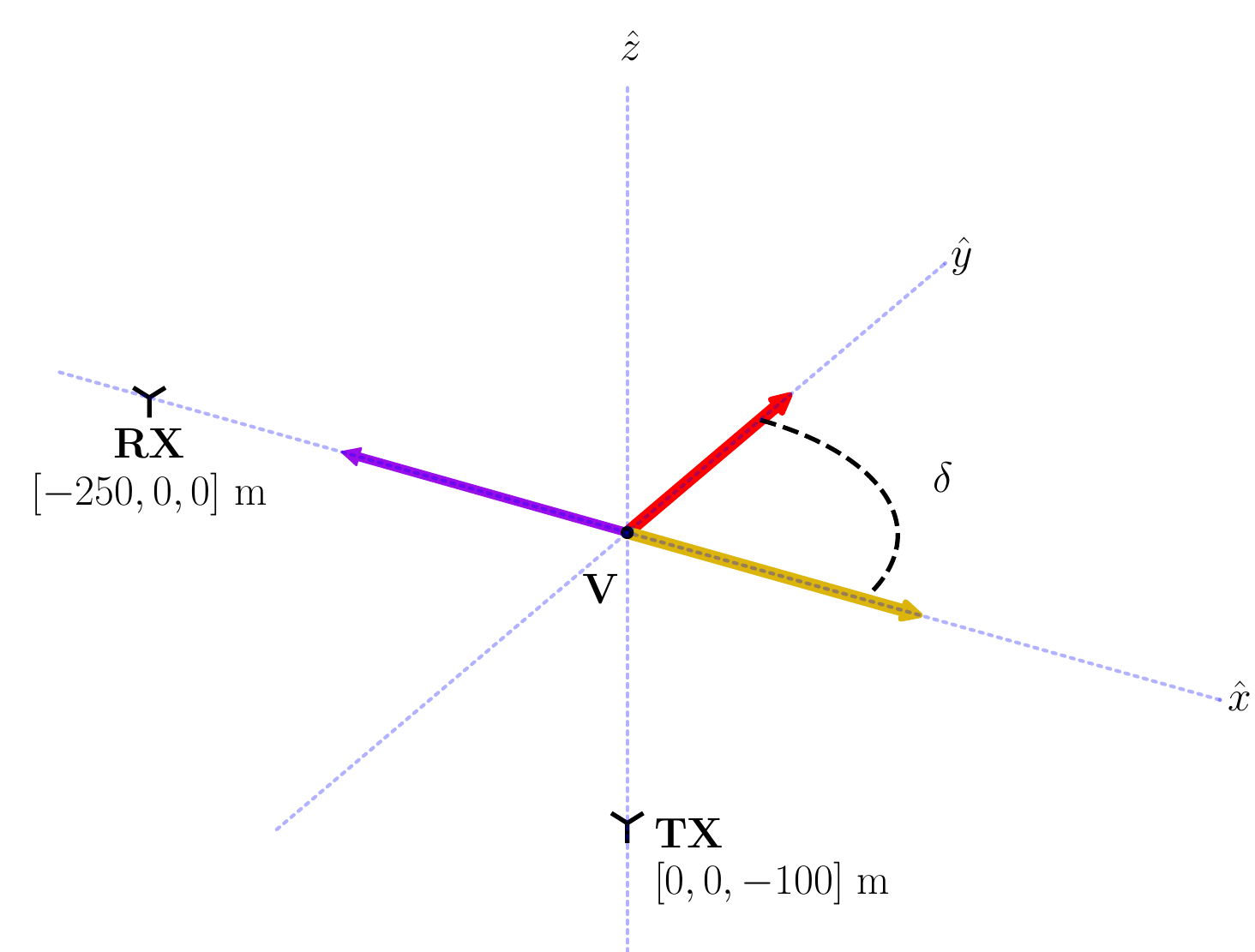}
         \caption{The cascade interaction point is set at $V(0,0,0)$. The cascade direction is indicated by the colored arrows. The cascade is rotated in the $xy$-plane through the angle $\delta$, defined by $\delta=0^\circ$ (yellow) for cascade propagation along the x-axis and $\delta=90^\circ$ (red) along the y-axis and $\delta=180^\circ$ (purple) along the $-\hat{x}$-axis. The receiver is located in the $xy$-plane at $Rx(-250,0,0)$~m. The transmitter is located at $z=-100$~m, directly below the cascade vertex at $Tx(0,0,-100)$~m.}
         \label{fig:geometry}
     \end{subfigure}
     \hfill
     \begin{subfigure}[b]{\linewidth}
         \centering
         \includegraphics[width=\textwidth]{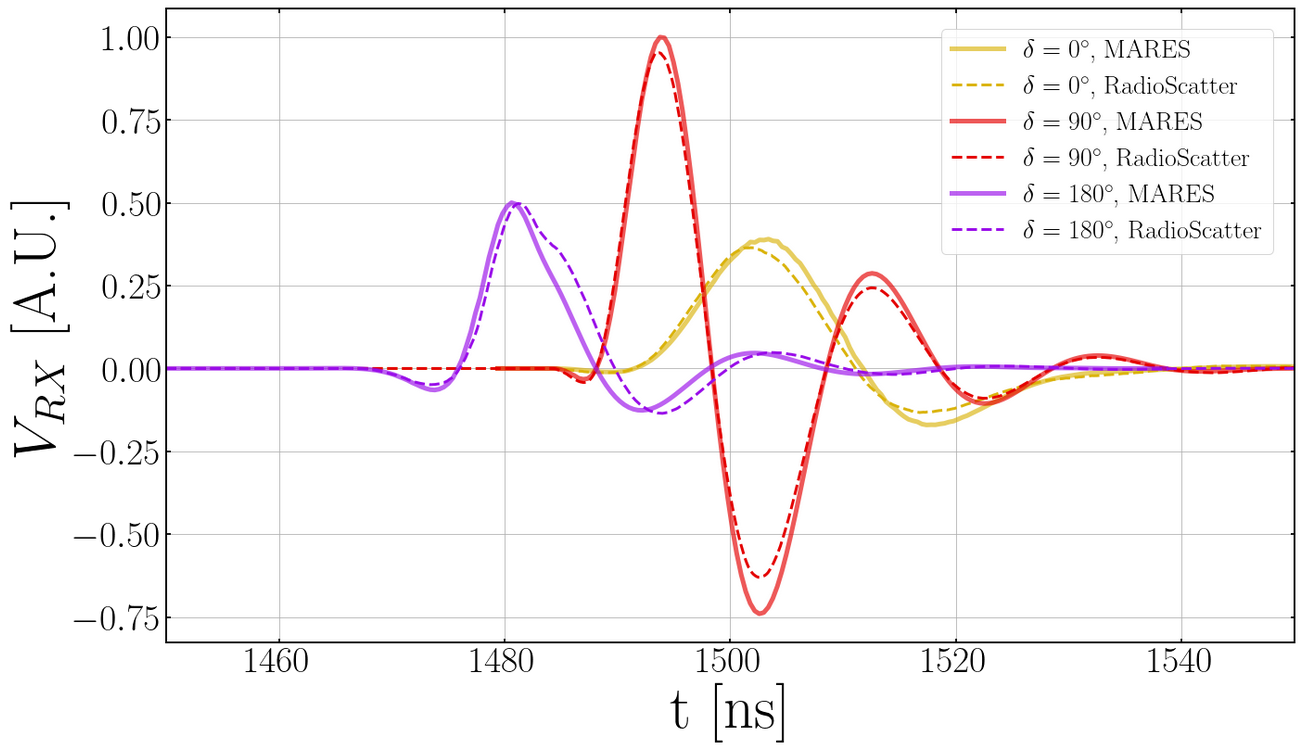}
         \caption{ \textsc{MARES} waveforms considering three events in the $xy$-plane at $\delta=0^\circ$ (yellow striped line) $\delta=90^\circ$ (full red line) and $\delta=180^\circ$ (purple dotted line). At $t = 0$ the cascade is initiated. The amplitude has been normalized to the highest voltage.}
         \label{fig:realistic_cascade_b}
     \end{subfigure}
     \hfill
     \caption{}
    \label{fig:realistic_cascade}
\end{figure}

As introduced in Sec.~\ref{sec:cascade}, the free-ionization electron lifetime determines the total number of charges available for scattering at any instant in time. If a cascade of energy $E_p$ starts at a $t = 0$, the amount of free electrons at any given instant can be described as the convolution of Eqs. (\ref{eq:Greisen}) and (\ref{eq:lifetime_factor}):
\begin{equation}
    N_{e^-,ion}(E_p, t, \tau) = N(E_p,t \cdot c \cdot \rho_{ice}) \circledast e^{-\frac{t}{\tau}} \Theta(t),
\end{equation}
where we make use of $c$ (the speed of the cascade front) and the density of the medium $\rho_{ice}$ to find the column density. This is shown in Fig.~\ref{fig:lifetime_10PeV}, where we observe that $\tau$ shapes the peak number of electrons available for scattering ($ \{N_{ion}(t,\tau)\}_\mathrm{max}$) and the moment in time when it happens, both important effects for the radar scatter signal. 

The effect on the plasma lifetime for a typical signal is shown in Fig.~\ref{fig:lifetime_waveforms}. It is seen that for short lifetimes, below approximately 30~ns, the pulse amplitude increases. For longer lifetimes, the amplitude is constant and maximal, as effectively the entire ionization trail is alive at a certain moment in time. Hence, for lifetimes longer than approximately 30~ns, the amplitude is no longer increasing, but the pulse duration is, and a clear exponential decay with its duration determined by the lifetime becomes apparent. 

\subsection{Realistic cascade simulations}   

The limited plasma lifetime and its relativistic propagation cause the scattering from different segments to interfere, and geometry-dependent signal features appear. This is investigated by considering different geometries for which relativistic propagation effects become apparent. To outline the effects of the relativistic propagation of the cascade, we show the simulated electric fields for three different radar scatter events with different geometries, presented in Fig.~\ref{fig:realistic_cascade}. 

The geometry under consideration is shown in Fig.~\ref{fig:geometry}, where we fix the interaction vertex to $(x,y,z)=(0,0,0)$~m, the receiver is located at $(x,y,z)=(-250,0,0)$~m and the transmitter is fixed 100~m below the interaction vertex at $(x,y,z)=(0,0,-100)$~m. 

It should be noted that the described signal features strongly depend on the transmitter location relative to the cascade. As such, the features described below provide intuition on the process, but are not general given that we fix our transmitter. The cascade direction is restricted to the $xy$ plane, varying the angle $\delta$, where for $\delta=0^\circ$ the cascade is moving away from the receiver in the $+\hat{x}$ direction, rotating clockwise to $+\hat{y}$, the direction defined by $\delta=90^\circ$. The cascade direction is indicated by the purple, red and yellow arrows. The same color scheme is used in Fig.~\ref{fig:realistic_cascade_b}, to indicate the corresponding waveforms for these geometries.

Fig.~\ref{fig:realistic_cascade_b} shows the simulated voltage at the receiver for a transmit frequency of 50~MHz, considering different cascade directions in the $xy$ plane. The computed electric field is converted into voltage using an antenna effective length of $h_{R}=1$~m. The results are shown for both \textsc{MARES} (solid lines) as well as \textsc{RadioScatter} (dashed lines) and all amplitudes are normalized to the peak pulse-height obtained from \textsc{MARES} considering the $\delta=90^\circ$ configuration.
 
Here, relativistic effects become apparent. The pulse is most compressed containing higher frequencies relative to the transmit frequency when the cascade is moving toward the receiver ($\delta=180^\circ$, purple dotted line), and stretched, containing lower frequencies, when it is moving away from the receiver ($\delta=0^\circ$, yellow striped line). The higher amplitude pulse is seen at ($\delta=90^\circ$, red full line). The obtained results (full lines) are compared to simulations performed using the \textsc{RadioScatter} code (dashed lines). The results agree very well on both pulse shape and amplitude, with differences found at the percent level for the presented geometries.

The overall more complicated three-body (transmitter-cascade-receiver) geometry demands for a more elaborate investigation to obtain a detailed explanation for the full phase space. This study is currently ongoing and will also allow for a more elaborate comparison between \textsc{MARES} and \textsc{RadioScatter}.

\subsection{The T-576 experiment}
To date, the T-576 experiment at SLAC provides the only positive observation of a radar scatter from a high-energy particle cascade~\cite{Prohira2020}. The target material in this case was not ice, but high density polyethylene (HDPE), where it is noted that HDPE has very similar properties to ice. This provides us with a benchmark to test our model, and allows providing an independent verification of the obtained experimental results. In the following, we use the parameters published in~\cite{Prohira2019b,Prohira2020}, summarized in Table~\ref{table:T576}. We deviate from our nominal value for the lifetime ($\tau_{nominal}=10$~ns), and use the value that was estimated in~\cite{Prohira2020}, based on the T576 observation, $\tau=3$~ns. Given the strong signal dependence on lifetime discussed in Fig.~\ref{fig:lifetime_waveforms}, we estimate its error at the ns level. The free charge collision frequency $\nu_c=100$~THz, as well as the mean electron ionization energy $E_{ion}=20$~eV are kept at their nominal values.  The transmitter and receiver antenna are modeled with their forward gains at $G_T= 12$~dBi, and $G_R=18$~dBi equal to the antennas used at T576 as reported in~\cite{Prohira2020}. Since, at T576, the antennas were directed at the expected cascade maximum, the directivity pattern given by their shape should not have a major effect on the simulated signal.

\begin{figure}[!h]
    \centering
    \includegraphics[width=\linewidth]%
    {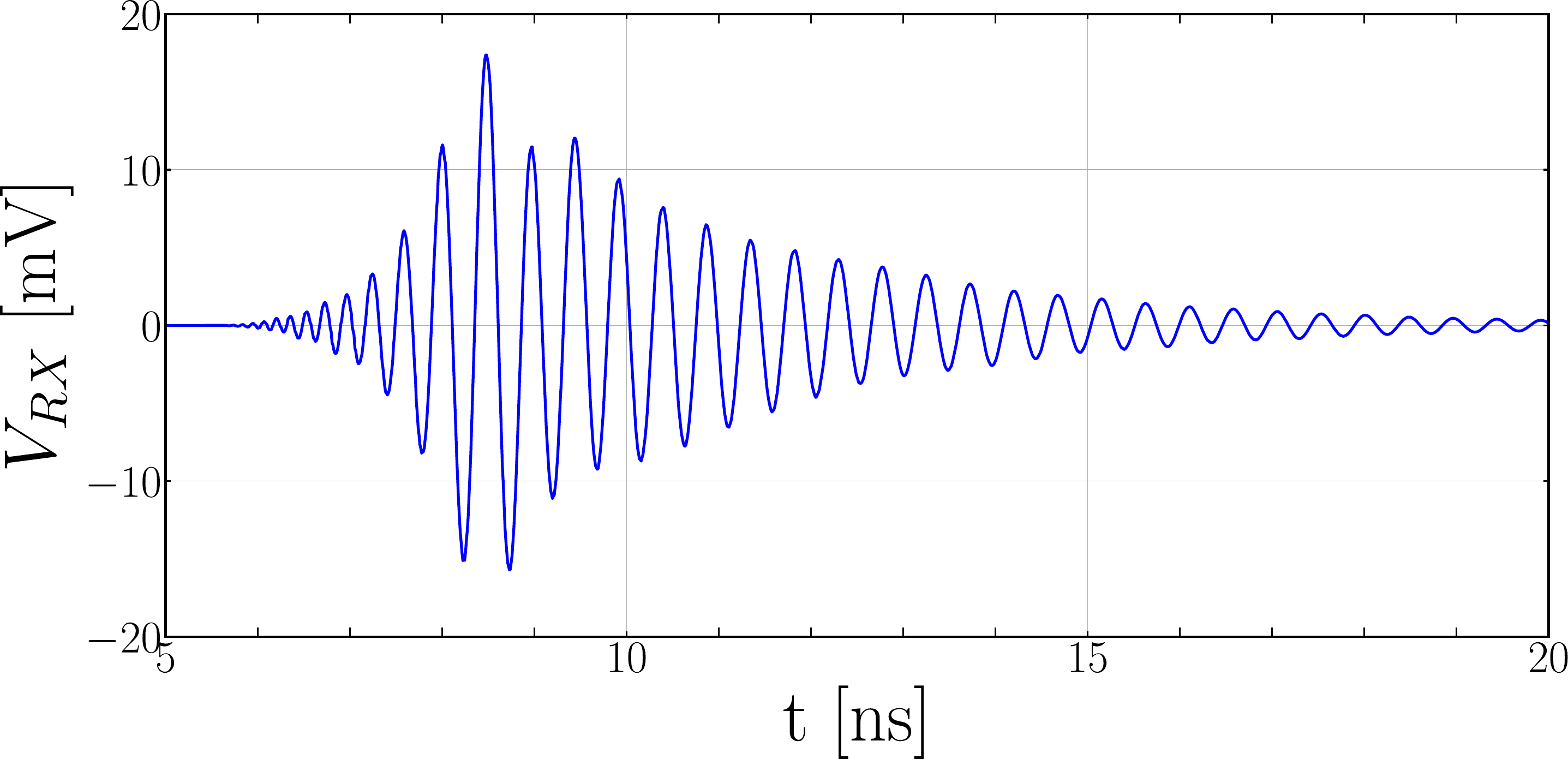}
    \caption{The predicted \textsc{MARES} signal of the T-576 experiment \cite{Prohira2020}}
    \label{fig:T576_voltage}
\end{figure}
\begin{table}[!h]
    \centering
    \begin{tabular}{|l|l|}
    \hline
       Parameters & Signal error estimate\\
       \hline
       Experimental configuration&\\
       \hline
       $E_p = 10~\mathrm{GeV}/\mathrm{e}^-$ &\\
       $N_p = 10^9~\mathrm{e}^-$ &\\
       $\vec{x}_t = (-3.8, 0, -0.3)~\mathrm{ m}$&\\
       $\vec{x}_r= (-3.877, 0, 0.3)~\mathrm{ m}$&\\ 
       $\vec{x}_c= (0,0,0)~\mathrm{m}$&\\
       $\vec{\beta}_c= (0,0,1)$&\\ 
       $\nu_T= 2.1~\mathrm{GHz}$ & $\sim 7$~dB (see Table 1 of ~\cite{Prohira2020}) \\
       $P_T= 50 ~\mathrm{W}$ &\\
       $G_T = 12~\mathrm{dB}$& \\
       $G_R = 18~\mathrm{dB}$& \\
       $n = 1.51$ &\\
       \hline
       \textsc{MARES} parameters& Signal error estimate\\
       \hline
       $E_{ion} = (20 \pm 2 )~\mathrm{eV} /\mathrm{e}^-$ & $O(10\%)$\\
       $\nu_c = (100 \pm 50)~\mathrm{THz}$ & $O(100\%)$ \\ 
       $\tau = (3 \pm 1)$~ns & $O(33\%)$ \\
       \hline
    \end{tabular}
    \caption{T576 parameters used as input for the \textsc{MARES} simulation, including error estimates. Error estimates on these parameters are taken from~\cite{Prohira2020} if provided and elaborated upon in the text.}
    \label{table:T576}
\end{table}

We performed a full \textsc{MARES} simulation using the framework outlined above, for which the result is shown in Fig.~\ref{fig:T576_voltage}. As we provide an estimate that matches reality well in pulse duration (using the 3~ns found in~\cite{Prohira2020}) frequency content and amplitude, it is important to note the errors on the different parameters available for the simulation. As such, an estimate on these errors is also given in Table~\ref{table:T576}. It follows that our estimate should be interpreted at the order of magnitude level. To conclude, for the given geometry, the observed field amplitude is at the same order of magnitude compared to the amplitude found in Fig.~3 of~\cite{Prohira2020}.   

\section{Conclusions}
We developed the \textsc{MARES} model to describe the radar echo from the ionization plasma left in the wake of a high-energy particle cascade developing in a dense medium like ice. The formalism is deterministic and allows for a computation of the scattering process, with a single simulation taking $O(\mathrm{s})$. 

As a first application, we find good agreement with the radar echo signal observed at the SLAC T-576 experiment. Furthermore, we studied the effect of the free ionization charge lifetime on the expected radar signal and introduce a radar scatter example that highlights the relationship between the geometry of the scatter and the signal features. Both examples show how a deterministic model like \textsc{MARES} can be used to quantify these observables.

The \textsc{MARES} code is currently available within the RET collaboration, and will be made public in the near future. 

\begin{acknowledgments}
We recognize support from The National Science Foundation under Grants No. 2012980, No. 2012989, No. 2306424, and No. 2019597 and the Office of Polar Programs, the Flemish Foundation for Scientific Research FWO-G085820N, the European Research Council under the European Unions Horizon 2020 research and innovation program (Grant Agreement No. 805486), the Belgian Funds for Scientific Research (FRS-FNRS), IOP, and the John D. and Catherine T. MacArthur Foundation. We would like to thank Carolina Huesca Santiago for producing Fig~\ref{fig:cascade_segment}.
\end{acknowledgments}

\appendix
\section{EQUATION OF MOTION}\label{app:eom}
For an incoming plane wave with angular frequency $\omega=2\pi\nu$ and wave number $k=2\pi/\lambda$ following the dispersion relation $\lambda \nu = c/n$ with $c$ the speed of light and $n$ the refractive index of the medium, the equation of motion for a single free charge in the plasma volume is properly modeled as a damped-driven oscillator: 
\begin{equation}
\label{eq: DDO_eq_motion}
    {\ddot{\vec{x}}}+\nu_c{\dot{\vec{x}}} + \omega_0^2\vec{x} = \vec{F}(t) = \frac{q E_e}{m} \cos({k x}-\omega t) \hat{p}. 
\end{equation}
The driving force is the the electromagnetic force induced by the transmitted radio wave, $\vec{F}(t) = q\vec{E}_e \cos (\omega t)$. The field is oscillating in the $\hat{p}$ direction given by the polarization of the incoming wave, and is driven by the electric field at the scattering location $\vec{E}_e(\vec{x},t)$. The solution for the free particle case ($\omega_0 = 0$) is found to be~\cite{TaylorClassical}
%
\begin{equation} \label{eq: DDO}
\begin{split}
\vec{x}_e 
    &=\frac{qE_e}{m} \left|\frac{1}{\omega^2 - i\omega\nu_c} \right| \cos({k x}-\omega t) \hat{p} \\
    &= \frac{qE_e}{m} |W| \cos({k x}-\omega t) \hat{p} \\
    & =  A_0 |W| \cos({k x}-\omega t) \hat{p}. \\
\end{split}
\end{equation}
The amplitude of oscillation is $|x_e| = A_0 |W|$, written in terms of the collisionless amplitude $A_0= qE_e m^{-1}$, where $q$ and $m$ respectively denote the charge and mass of the oscillator. The amplitude damping factor, $W$, is a function of the characteristic frequencies in the system and is defined as
\begin{equation}
    W=\left(\frac{1}{\omega^2 - i\omega \nu_c} \right). 
\end{equation}
Having solved the equation of motion for the scatter, we obtain the scattered electric field at an arbitrary observer location through the standard field equations, 
\begin{equation} \label{eq: field-e-R}
\begin{split}
\vec{E}_{e,R} 
    &= \left[ \frac{q}{(4 \pi \epsilon_0 ) c^2} \right] \frac{1}{R_R}
        [\hat{n} \times (\hat{n} \times \hat{\ddot{x}}_e) ]  \\
    &= \left[ \frac{q}{(4 \pi \epsilon_0 ) c^2} \right] \frac{1}{R_R}
       [\hat{n} \times (\hat{n} \times \omega^2\hat{x}_e) ]  \\
    &= \left[ \frac{q}{(4 \pi \epsilon_0 ) c^2} \right] \frac{\omega^2}{R_R}
      |x_e \sin(\theta)|\;\;\hat{p} \\
    &= \left[ \frac{q}{(4 \pi \epsilon_0 ) c^2} \right] \frac{\omega^2}{R_R}
      \left|\frac{qE_e}{m} W \right| | \sin(\theta)|\;\;\hat{p} \\
    &= \left[ r_e \right] E_e \frac{\omega^2 |W|}{R_R}
      | \sin(\theta)|\;\;\hat{p} \\
    &=  E_e \frac{\omega^2 |W|}{R_R} \left( \sqrt{\frac{8\pi}{3}} r_e \right) \left( \sqrt{\frac{3}{8\pi}} | \sin(\theta)| \right) \;\;\hat{p} \\
  &=  E_e \frac{\omega^2 |W|}{R_R} \sqrt{ \frac{\sigma_{Th}G_{Hz}}{4\pi}} \;\;\hat{p}
\end{split}
\end{equation} 

 The final expression is written in terms of the Thomson scattering cross section  $\sigma_{Th} = \frac{8\pi}{3}r_e^2$, for which all constants sum to the classical electron radius $r_e~\sim~2.8~\times~10^{-15}$~m, and the Hertzian dipole gain factor $G_{Hz}~=\frac{3}{2}\sin^2(\theta)$. Here $\hat{n}$ denotes the unit vector pointing from the charge to the receiver and $\theta$ is here the angle between $\hat{p}$ and $\hat{n}$, originating from the cross product in the numerator of Eq.~(\ref{eq: field-e-R}).

We can insert the expression of $|E_e|$ found in Eq.~(\ref{eq: E_field_at_e}) in Eq.~(\ref{eq: field-e-R}) and compare the result to Eq.~(\ref{eq: radar_field}).

This now allows us to identify the radar scattering cross section of the damped free electron:
\begin{equation}\label{eq: RCSe}
    \sigma_{RCS,e} = (\omega^2W)^2 \sigma_{Th} G_{Hz}.
\end{equation}

\section{FRAMES OF REFERENCE}
\subsection{\label{app:reference} The bistatic scattering frame}

First, we define the bistatic scattering frame by placing the origin at the transmitter (TX), the $\vec{e}_{rx}$ unit vector points from TX to the receiver (RX), and $\vec{e}_z$ is the vertical. 

We now define
\begin{equation}
  \vec{e}_y=\vec{e}_z\times\vec{e}_{rx} ; \ 
  \vec{e}_x=\vec{e}_y\times\vec{e}_z,
\end{equation}
such that $\{\vec{e}_x,\vec{e}_y,\vec{e}_z\}$ defines a right-handed orthogonal basis. In this frame, the cascade is defined by the position of the interaction vertex $\vec{x}_{c}$, and the cascade direction  $\vec{\beta}_{CS}$. This frame allows to directly evaluate the bistatic radar scattering equation for our segments.

\subsection{The cascade frame}
The cascade frame is defined by the cascade direction $\vec{e}_{\beta_{CS}}$ and the direction vector pointing from the cascade to the transmitter, $\vec{d}_{tx} = $ $\vec{x}_{c}- \vec{x}_{t}$, with unit vector $\vec{e}_{d_{tx}}$. The cascade frame is defined by $\{\vec{e}_L, \vec{e}_R, \vec{e}_n\}$. Here,
\begin{equation}
  \vec{e}_L = \vec{e}_{\beta_{CS}}; \ 
  \vec{e}_n = \vec{e}_{L} \times \vec{e}_{d_{tx}}; \ 
  \vec{e}_R = \vec{e}_n \times \vec{e}_{L}.
\end{equation}
This frame allows for a universal cascade parametrization along its longitudinal direction $\vec{e}_L$, and radial direction $\vec{e}_R$.

\subsection{The plane of incidence frame}
The plane of incidence frame is defined as $\{\vec{e}_U, \vec{e}_V, \vec{e}_W\}$ and is constructed similarly as the cascade frame, 
\begin{equation}
  \vec{e}_U=\vec{e}_{d_{tx}}; \ 
  \vec{e}_V=\vec{e}_n \times \vec{e}_{d_{tx}}; \ 
 \vec{e}_W=\vec{e_n}.
\end{equation}
The line of sight from the transmitter to the cascade, and thus the direction of propagation of the transmitted radio wave is defined by $\vec{e}_U$. The polarization of the emitted radio wave is now naturally found in the $\{ \vec{e}_V,\vec{e}_W\}$ plane.

 Note that this frame is orthogonal only in the plane wave approximation, which is valid under the condition that the effective length of the cascade in ice, $L$, is small compared to the distance to the antennas $R_{T}, R_{R}$, such that the incidence angle along the cascade is constant to good approximation.
 
\section{CONVERGENCE \label{app:convergence}}
\begin{figure}[!ht]
     \centering
      \begin{subfigure}[b]{\linewidth}
          \centering
     \includegraphics[width=\linewidth]{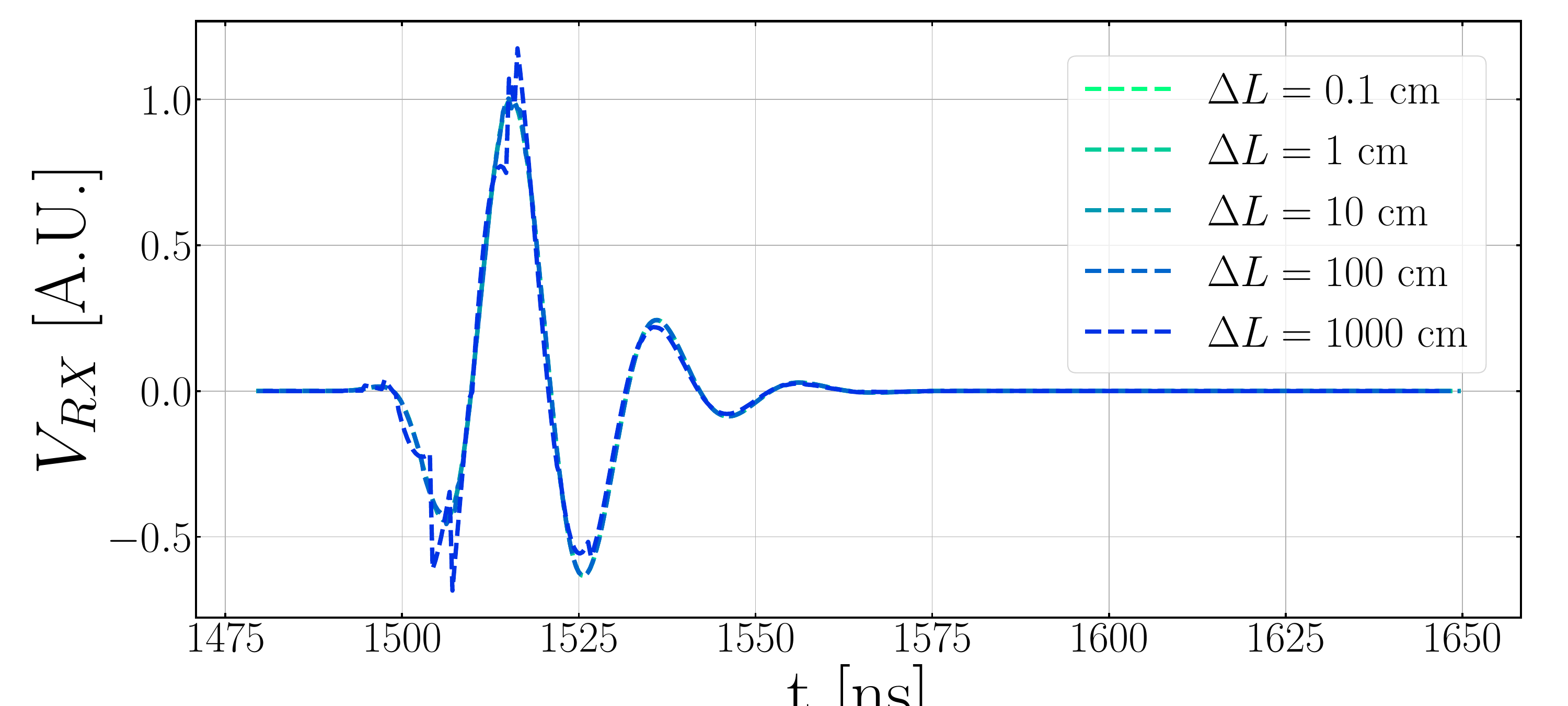}
     \caption{Convergence of the step size parameter $\Delta L$, fixing $\Delta r$ to its nominal value, $\Delta r=1$~mm.}
      \label{fig:Convergence_dL}
      \end{subfigure}
      \begin{subfigure}[b]{\linewidth}
          \centering
    \includegraphics[width=\linewidth]{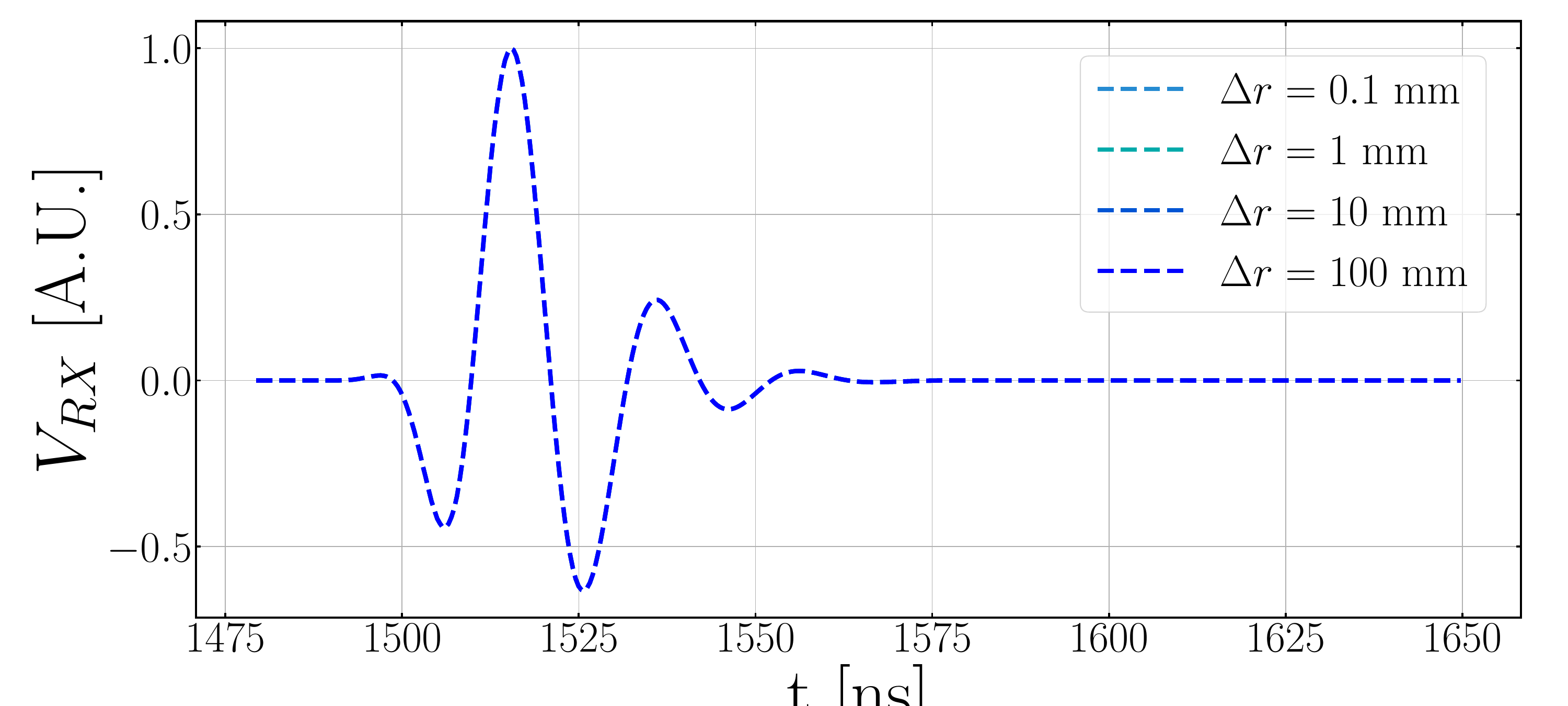}
    \caption{Convergence of the step size parameter $\Delta r$, fixing $\Delta L$ to its nominal value, $\Delta L=1$~cm. }
    \label{fig:Convergence_dr}
    \end{subfigure}
    \caption{Convergence over the segmentation step size parameters $\Delta L$ and $\Delta r$ considering the geometry outlined in Fig.~\ref{fig:geometry}, using $\delta=90^\circ$, and $\theta=90^\circ$. The pulse peak amplitude is normalized to the amplitude found using the nominal values $\Delta L=1$~cm and $\Delta r=1$~mm.}
\end{figure}

To check for convergence of the \textsc{MARES} model over its segmentation steps, $\Delta L$ and $\Delta r$, we consider the geometry outlined in Fig.~\ref{fig:geometry}, using $\delta=90^\circ$, and $\theta=90^\circ$, for which the projected cascade size towards the observer is maximized. In Fig.~\ref{fig:Convergence_dL}, we fix $\Delta r$ to its nominal value $\Delta r=1$~mm, and vary $\Delta L$, where in Fig.~\ref{fig:Convergence_dr} we fix $\Delta L$ to its nominal value $\Delta L=1$~cm, and vary $\Delta r$. It follows that the results converge for step sizes roughly a factor of 10 above the nominal values used for this work. However, given that this only considers a single geometry, we recommend using the nominal step size when running the simulation. 

\section{FREE ELECTRON LIFETIME \label{app:fel}}

\begin{figure}[!h]
    \centering
    \includegraphics[width=\linewidth]{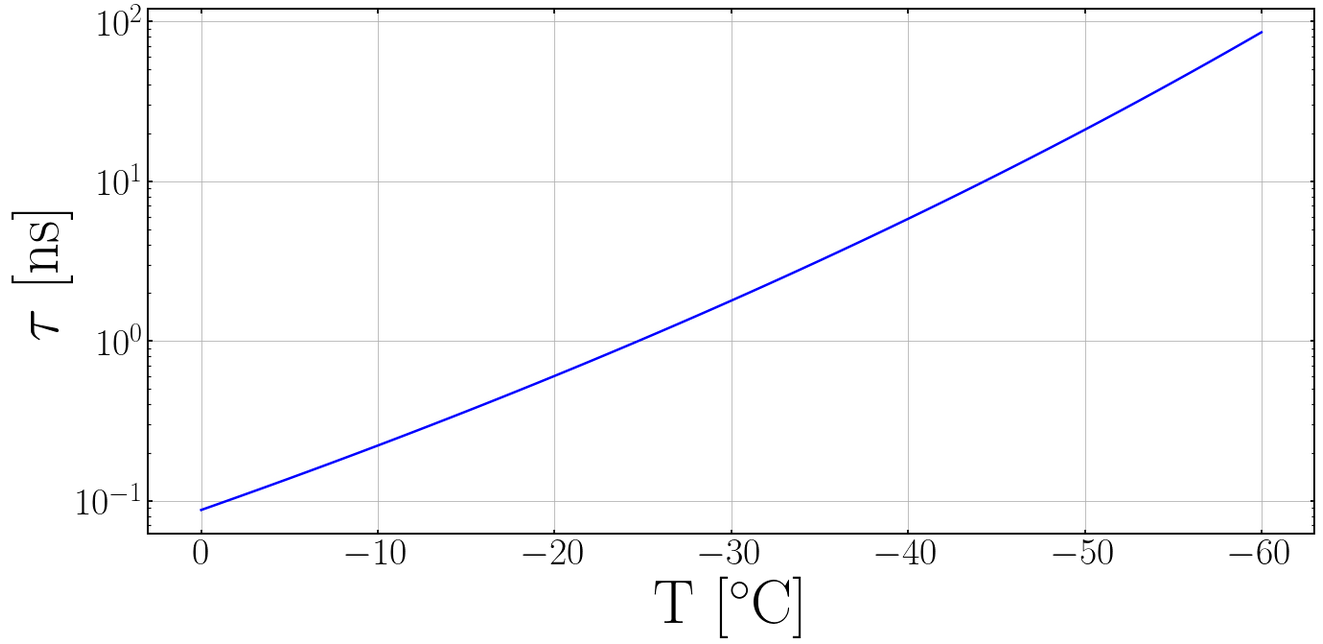}
    \caption{The free electron lifetime as function of temperature.}
    \label{fig:fel}
\end{figure}

The total number of free ionization electrons available for scattering is well described through a standard decay $N(t,\tau)=N(t=0,\tau)e^{-t/\tau}$, assuming the ionization occurs at the time $t=0$. Here, $\tau$ denotes the free electron lifetime.

As such, we are interested in obtaining an accurate value, or range of values, for the free electron lifetime, $\tau$, for ice. Over the past few decades, several measurements have been made that allow to determine $\tau$ for ice. We are interested in the temperature range of ice typically found in polar regions, $T=[-60\si{\celsius}, 0\si{\celsius}]$~\cite{Buford2002,MacGregor2015}. Therefore, following a recent overview work~\cite{Khamzin2010}, we base ourselves on results obtained in 1952 by Auty and Cole~\cite{AutyCole1952}, and related measurements by De Haas \textit{et. al.}, in 1983~\cite{DeHaas1983}.  

Auty and Cole took measurements of the dielectric constant of ice to obtain the dielectric relaxation time $\tau_c$ (not to be confused with the free electron lifetime), 
\begin{equation}
    \tau_c = A e^{\frac{B}{k_B T}} 
\end{equation}
with A = $5.3\times10^{-16}$~s, $B=0.575$~eV, and $k_B=8.617 \times 10^{-5}\;\mathrm{eV \;K^{-1}}$, where the latter is Boltzmann's constant. \\

These results were subsequently used by De Haas \textit{et. al.}, who performed a series of measurements to obtain the free electron lifetime in ice as function of temperature. This was done by directing x-ray or 3~MeV electron bunches into a block of ice, after which the ice conductivity was obtained as a function of time~\cite{DeHaas1983}.

In Fig. 3 of~\cite{DeHaas1983}, the electron trapping rate in ice, $f_r=1/\tau \mathrm{[s^{-1}]}$, is presented as a function of reciprocal temperature $10^3/T\;\mathrm{[K^{-1}]}$ and fitted using the Arrhenius equation to obtain
\begin{equation} \label{eq: lifetime_vs_T}
    \tau = \tau_c/C= (A/C) e^{\frac{B}{k_B T}}
\end{equation}
with $C$ a scale factor reported in the caption of Fig.~3 of ~\cite{DeHaas1983}, equal to $C=1/(2.5\times10^5)$, and $\tau_C$ is taken from the Auty and Cole measurements. This allows us to calculate the free electron lifetime over our temperature range of interest, $T=[-60\si{\celsius}, 0\si{\celsius}]$, as shown in Fig.~\ref{fig:fel}. It follows, the lifetime ranges from roughly 0.1~ns at $0\si{\celsius}$ up to $100$~ns at $-60\si{\celsius}$. 

The temperature of polar ice has been well studied, and its variation with factors like depth, weather or the season are well characterized for South Pole and Greenland, both for the top layers of the ice/snow (firn) \cite{Brandt1997} \cite{Giese2015} and the deeper ice (bulk) \cite{Buford2002} \cite{MacGregor2015}. If the polar ice follows the Arrhenius equation, Eq.~(\ref{eq: lifetime_vs_T}), then the free electron lifetime, a key but, in principle, free parameter of the model can be well understood and controlled. However, there are other properties of the polar ice, like the concentration of impurities, whose role on the free electron lifetime is not well characterized.

\bibliographystyle{JHEP}
\bibliography{references}

\end{document}

%% file: author_list.tex
\author{E. Huesca Santiago}
 \email{enrique.huesca.santiago@vub.be}
 \affiliation{Vrije  Universiteit  Brussel, Dienst ELEM, IIHE,  1050 Brussels,  Belgium}
\author{K.D. de Vries}%
 \email{krijn.de.vries@vub.be}
\affiliation{Vrije  Universiteit  Brussel, Dienst ELEM, IIHE,  1050 Brussels,  Belgium}

\author{P.~Allison}
\affiliation{Department of Physics, Center for Cosmology and AstroParticle Physics (CCAPP), The Ohio State University, Columbus Ohio 43210, USA}
\author{J.~Beatty}
\affiliation{Department of Physics, Center for Cosmology and AstroParticle Physics (CCAPP), The Ohio State University, Columbus Ohio 43210, USA}
\author{D.~Besson}
\affiliation{University of Kansas, Lawrence, Kansas 66045, USA}
\author{A.~Connolly}
\affiliation{Department of Physics, Center for Cosmology and AstroParticle Physics (CCAPP), The Ohio State University, Columbus Ohio 43210, USA}
\author{A.~Cummings}
\affiliation{Departments of Physics and Astronomy \& Astrophysics, Institute for Gravitation and the Cosmos,   Pennsylvania State University, University Park, Pennsylvania 16802, USA}
\author{C.~Deaconu}
\affiliation{Enrico Fermi Institute, Kavli Institute for Cosmological Physics, Department of Physics, University of Chicago, Chicago, Illinois 60637, USA}
\author{S.~De~Kockere}
\affiliation{Vrije  Universiteit  Brussel, Dienst ELEM, IIHE,  1050 Brussels,  Belgium}
\author{D.~Frikken}
\affiliation{Department of Physics, Center for Cosmology and AstroParticle Physics (CCAPP), The Ohio State University, Columbus Ohio 43210, USA}
\author{C.~Hast}
\affiliation{SLAC National Accelerator Laboratory, Menlo Park, California 94025, USA}
\author{C.-Y.~Kuo}
\affiliation{National Taiwan University, Taipei, Taiwan}
\author{A.~Kyriacou}
\affiliation{University of Kansas, Lawrence, Kansas 66045, USA}
\author{U.A.~Latif}
\affiliation{Vrije  Universiteit  Brussel, Dienst ELEM, IIHE,  1050 Brussels,  Belgium}
\author{I.~Loudon}
\affiliation{Department of Astrophysics/IMAPP, Radboud University, P.O. Box 9010, 6500 GL Nijmegen, The Netherlands}
\author{V.~Lukic}
\affiliation{Vrije  Universiteit  Brussel, Dienst ELEM, IIHE,  1050 Brussels,  Belgium}
\author{C.~McLennan}
\affiliation{University of Kansas, Lawrence, Kansas 66045, USA}
\author{K.~Mulrey}
\affiliation{Department of Astrophysics/IMAPP, Radboud University, P.O. Box 9010, 6500 GL Nijmegen, The Netherlands}
\author{J.~Nam}
\affiliation{National Taiwan University, Taipei, Taiwan}
\author{K.~Nivedita}
\affiliation{Department of Astrophysics/IMAPP, Radboud University, P.O. Box 9010, 6500 GL Nijmegen, The Netherlands}
\author{A.~Nozdrina}
\affiliation{University of Kansas, Lawrence, Kansas 66045, USA}
\author{E.~Oberla}
\affiliation{Enrico Fermi Institute, Kavli Institute for Cosmological Physics, Department of Physics, University of Chicago, Chicago, Illinois 60637, USA}
\author{S. Prohira}
\affiliation{University of Kansas, Lawrence, Kansas 66045, USA}
\author{J.P.~Ralston}
\affiliation{University of Kansas, Lawrence, Kansas 66045, USA}
\author{M.F.H.~Seikh}
\affiliation{University of Kansas, Lawrence, Kansas 66045, USA}
\author{R.S.~Stanley}
\affiliation{Vrije  Universiteit  Brussel, Dienst ELEM, IIHE,  1050 Brussels,  Belgium}
\author{J.~Stoffels}
\affiliation{Vrije  Universiteit  Brussel, Dienst ELEM, IIHE,  1050 Brussels,  Belgium}
\author{S.~Toscano}
\affiliation{Universit\'{e} Libre de Bruxelles, 1050 Brussels, Belgium}
\author{D.~Van~den~Broeck}
\affiliation{Vrije  Universiteit  Brussel, Dienst ELEM, IIHE,  1050 Brussels,  Belgium}
\affiliation{Vrije  Universiteit  Brussel, Astrophysical Institute, 1050 Brussels,  Belgium}
\author{N.~van~Eijndhoven}
\affiliation{Vrije  Universiteit  Brussel, Dienst ELEM, IIHE,  1050 Brussels,  Belgium}
\author{S.~Wissel}
\affiliation{Departments of Physics and Astronomy \& Astrophysics, Institute for Gravitation and the Cosmos,   Pennsylvania State University, University Park, Pennsylvania 16802, USA}
\collaboration{Radar Echo Telescope Collaboration}